\documentclass[%
 reprint,
 amsmath,amssymb,
 aps,
floatfix,
]{revtex4-2}

\usepackage{graphicx}
\usepackage{dcolumn}
\usepackage{bm}

\usepackage[caption=false]{subfig}
\usepackage[normalem]{ulem} 

\usepackage{siunitx} 
\begin{document}

\title{Laser-induced damage threshold of periodic plasmonic metasurfaces}
\author{Sebastian Beer}
\affiliation{Friedrich Schiller University Jena, Institute of Applied Physics, Abbe Center of Photonics, Albert-Einstein-Str. 15, 07745 Jena, Germany}

\author{Amit Shaha Surja}
\affiliation{Friedrich Schiller University Jena, Institute of Applied Physics, Abbe Center of Photonics, Albert-Einstein-Str. 15, 07745 Jena, Germany}

\author{Jeetendra Gour}%
\affiliation{Friedrich Schiller University Jena, Institute of Applied Physics, Abbe Center of Photonics, Albert-Einstein-Str. 15, 07745 Jena, Germany}

\author{Alessandro Alberucci}
\affiliation{Friedrich Schiller University Jena, Institute of Applied Physics, Abbe Center of Photonics, Albert-Einstein-Str. 15, 07745 Jena, Germany}

\author{Uwe Zeitner}
\author{Stefan Nolte}
\affiliation{Friedrich Schiller University Jena, Institute of Applied Physics, Abbe Center of Photonics, Albert-Einstein-Str. 15, 07745 Jena, Germany}
\affiliation{Fraunhofer Institute for Applied Optics and Precision Engineering IOF, Albert-Einstein-Str. 7, 07745 Jena, Germany}

\date{\today}

\begin{abstract}
Plasmonic metasurfaces enable strong nonlinear optical responses due to sub-wavelength field confinement, yet their performance is fundamentally limited by low laser-induced damage thresholds (LIDT). 
Here, we present a comprehensive experimental and numerical study of the interplay between surface lattice resonances (SLRs), second-harmonic generation (SHG), and LIDT in periodic gold nanobar metasurfaces. 
Angle-resolved measurements reveal that excitation of SLRs at the fundamental wavelength strongly enhances SHG efficiency while simultaneously reducing the LIDT. 
In contrast, excitation of an SLR at the second-harmonic wavelength does not affect the damage threshold, allowing the SHG signal to be maximized.
Our findings establish a fundamental trade-off between nonlinear efficiency and optical robustness and provide design guidelines for optimizing periodic plasmonic metasurfaces for high-power nonlinear applications.
\end{abstract}


\maketitle

\section{Introduction}
Plasmonics has attracted sustained attention due to its ability to enhance light–matter interactions through sub-wavelength confinement of electromagnetic fields. 
In particular, the excitation of localized surface plasmon resonances (LSPRs) in metallic nanoparticles leads to strongly driven collective electron oscillations, resulting in pronounced near-field enhancement. 
When nanoparticles are periodically arranged to form a plasmonic metasurface, coherent coupling between the LSPRs of individual particels and the diffraction order of the lattice can give rise to a hybrid plasmonic–photonic mode known as surface lattice resonance (SLR) \cite{Augui2008, Kravets2018}.
The resulting collective resonance has much narrower spectral features than those of the isolated nanoparticles \cite{Humphrey2014}.
In suitably designed plasmonic arrays, SLRs can also provide stronger local field enhancements than LSPRs \cite{Bin-Alam2021}.

These properties make plasmonic metasurfaces attractive for nonlinear optical processes, such as second- \cite{Czaplicki2016,Michaeli2017,Hooper2019,Stolt2022,Beer2022}, third-harmonic generation \cite{Shen2020, Gour2022}, four-wave mixing \cite{Chakraborty2023}, saturable absorption \cite{Xie2024}, or high harmonic generation \cite{Kim2008, Sivis2013, Vampa2017}.
However, due to the intrinsic optical properties of commonly used metals, the field intensities required to drive significant optical interactions easily surpass the modification threshold of the plasmonic nanostructures, thereby compromising both the performance and durability of plasmonic devices \cite{Kern2015,Krasnok2018}.
Consequently, understanding and optimizing the critical fluence  above which the optical response is permanently altered, also known as laser induced damage threshold (LIDT), is essential for advancing plasmonic devices.


Resonantly excited plasmonic nanostructures can exhibit a reduced LIDT compared to unstructured metal films \cite{Kern2011}.
Three major categories of damage mechanisms have been identified \cite{Spellauge2025}: thermal, thermomechanical, and non-thermal.
Thermal damage originates from optical absorption, whereby the absorbed energy is initially deposited in the electron system and subsequently transferred to the lattice through electron–phonon coupling, leading to a rapid increase in lattice temperature \cite{Hohlfeld2000, Link2003}.
Under ultrashort pulse excitation, energy deposition and lattice heating are temporally separated.
Laser-induced heating can reshape plasmonic nanoparticles into more spherical structures through partial or complete melting \cite{Link1999,Link2000,Wang2021}, or thermally-activated surface diffusion below the complete melting temperature \cite{Inasawa2005,Taylor2014}.
At higher excitation fluences, fragmentation into smaller particles can occur \cite{Link1999,Link2000}. 
Thermomechanical damage can occur when lattice heating takes place faster than mechanical stress can relax. 
The resulting pressure buildup and subsequent stress relaxation can cause structural disintegration, including spallation or cavitation \cite{Zhigilei1998,Ihm2019}.
Non-thermal, field-driven processes can induce highly localized material removal at plasmonic hotspots \cite{Plech2006, Shi2017}.
In addition, nanoparticles can detach from the substrate \cite{Nagy2020}. 

Depending on the dominant damage mechanism, different strategies can be employed to increase the LIDT of plasmonic nanostructures.
Substrates with high thermal conductivity (such as sapphire, silicon or diamond) can improve heat dissipation and thereby increase the LIDT under thermally limited conditions \cite{Morshed2020,Tsibidis2022-10}. 
Increasing the thermal mass of plasmonic nanostructures, for example by increasing their thickness, can reduce thermally induced damage \cite{Shi2018}.
Alternatively, materials such as tungsten, chromium, and titanium nitride can provide higher laser damage resistance than conventional plasmonic metals such as gold and silver \cite{Morshed2019,Gui2016}.
Dielectric overcoatings can improve the damage resistance of metallic nanostructures and stabilize their nonlinear optical response \cite{Shi2018,Albrecht2017}.

Based on a comparison of unstructured gold films with non-resonant and resonant gold nanostructures, \citeauthor{Kern2015}~\cite{Kern2015} found that stronger spectral overlap between the plasmonic resonance and the excitation laser increases the susceptibility to laser-induced damage.
Consequently, accounting for thermal damage, the maximum achievable harmonic yield may occur off-resonance \cite{Pfullmann2013}.
This work focuses on the influence of SLR on the LIDT of plasmonic metasurfaces consisting of periodically arranged gold nanobars.
Although more complex nanostructure geometries may achieve higher nonlinear conversion efficiencies, simple gold nanobars are employed to focus on the influence of SLRs on the LIDT.
By varying the optical angle of incidence, different SLR of the metasurface are excited and their impact on the LIDT is investigated.
Furthermore, although limitations of plasmonic nanoparticles for frequency conversion application have been discussed \cite{Kern2015}, the interplay between SLR-enhanced SHG and the LIDT has not yet been systematically investigated.
Comparing the LIDT with the SHG efficiency provides further insight into a trade-offs between nonlinear enhancement and optical damage, helping to identify excitation conditions that maximize the achievable SHG output without inducing irreversible modifications.

\section{Sample Design}
\begin{figure}[htbp]
	\centering
	\includegraphics[width=.4\textwidth]{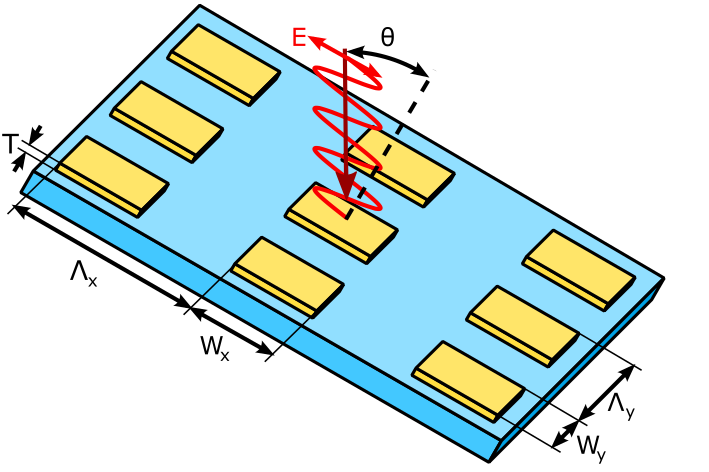}
	\caption{
		Schematic of the metasurface.
        The dimensions of the elemental NBs are $W_x=\SI{400}{\nano\meter}$, $W_y=\SI{300}{\nano\meter}$, and $T = \SI{50}{\nano\meter}$.
        The periodicities are $\Lambda_x = \SI{680}{\nano\meter}$ and $\Lambda_y = \SI{500}{\nano\meter}$.
        The $3\times\SI{3}{\milli\meter\squared}$ large metasurface is fabricated on a \SI{1}{\milli\meter}-thick fused silica substrate.
        }
	\label{Fig:SampleDesign}
\end{figure}

The investigated metasurface consists of periodically arranged gold nanobars (NBs) in a rectangular lattice on a fused silica substrate, as illustrated in Fig.~\ref{Fig:SampleDesign}.
The metasurfaces were fabricated using a standard metal lift-off process and electron-beam lithography for patterning.
In previous works we already investigated the linear response, SHG \cite{Beer2022, Beer2025}, and THG \cite{Gour2022} of these metasurfaces depending on $\Lambda_x$.

\section{Linear response}
\begin{figure}[htbp]
	\centering
	\includegraphics[width=.4\textwidth]{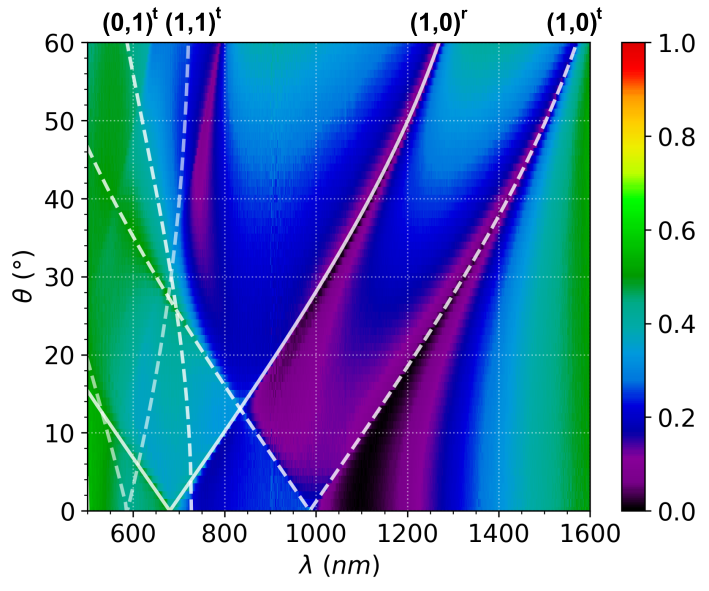}
	\caption{   
        Measured transmission spectra in dependence of the angle of incident ($\theta$).
        Superposed lines show the RA conditions related to the dedicated diffraction orders,
        solid lines for the reflection grating case and dashed lines for the transmission grating case.
        }
	\label{Fig:Linear}
\end{figure}

Figure~\ref{Fig:Linear} shows the transmission spectra of the metasurface in dependence of the angle of incidence ($\theta$).
Sharp SLR related discontinuities in the optical response are visible, whose spectral position changes with $\theta$.
The metasurface is tilted along its $y$-axis, thus the linear polarized light is TM/p polarized when the electric field vector is oriented along the $x$-direction of the metasurfaces, as illustrated in Fig.~\ref{Fig:SampleDesign}.
Close to the spectral position of the SLR, the diffracted light from the metasurface propagates parallel to the surface \cite{Dey2024, Khlopin2017}, this condition is known as Rayleigh anomaly (RA) \cite{Rayleigh1907}.
The spectral position of the RA condition is represented as superimposed lines in Fig.~\ref{Fig:Linear}.
The incident angles required to satisfy the RA condition ($\theta_{RA}$) are determined by momentum conservation for a two-dimensional lattice in the limit where the diffracted order propagates parallel to the metasurface plane:
\begin{align}
	\sin{\theta_{RA}} = -\frac{p\lambda}{n_1\Lambda_x} \pm \sqrt{\left( \frac{n_{(1,2)}}{n_1}\right)^2 - \left(\frac{q\lambda}{n_1\Lambda_y}\right)^2} \label{eq:theta_RA}.
\end{align}
The integers $p$ and $q$ denote the diffraction orders along the grating in the $x$-direction and $y$-direction, respectively.
The refractive index $n_{(1,2)}$ depends on the medium in which the diffracted light propagates.
The asymmetric refractive-index environment of the metasurface leads to two cases \cite{Auguie2010}: the reflection grating case where the diffracted light propagates in air ($n_{(1,2)} = n_1$) and the transmission grating case where it propagates in the substrate ($n_{(1,2)}= n_2$).
The diffraction orders are labeled as $(p,q)^r$ or $(p,q)^t$ for the reflection and transmission grating case, respectively.

The broad transmission minium at $\lambda \approx \SI{1150}{\nano\meter}$ is the hybrid mode \cite{Khlopin2017}, between the LSPR and the $(0,1)^t$ diffraction order.
In contrast, the SLR is another hybrid mode with narrower spectral features, higher quality factors, and stronger near-field enhancement \cite{Pan2023}.

\section{Second-harmonic generation}
An ultrafast laser featuring a pulse duration of \SI{270}{\femto \second}, a central wavelength of \SI{1030}{\nano \meter}, a repetition rate of \SI{200}{\kilo \hertz}, and a beam waist of $\omega_0 = \SI[separate-uncertainty=true]{61.5 \pm 0.6}{\micro\meter}$ in the focus of a \SI{400}{\milli\meter} lens is used to excite the metasurface.
Only the second-harmonic (SH) emission from the metasurface in direct transmission ($0^{th}$-diffraction order) is imaged onto a camera.
The fundamental-frequency (FF) pump light and environmental background are suppressed using optical filters, such that no signal is detected in the absence of a sample.
After subtracting the background level, the sum of the pixel values of the detected spot divided by the exposure time is proportional to the average optical power of the SH ($P_{SH}$), which has been calibrated to a power meter.

\subsection{Angle-dependent SHG response}
\begin{figure}[tbhp]
	\centering
	\includegraphics[width=.4\textwidth]{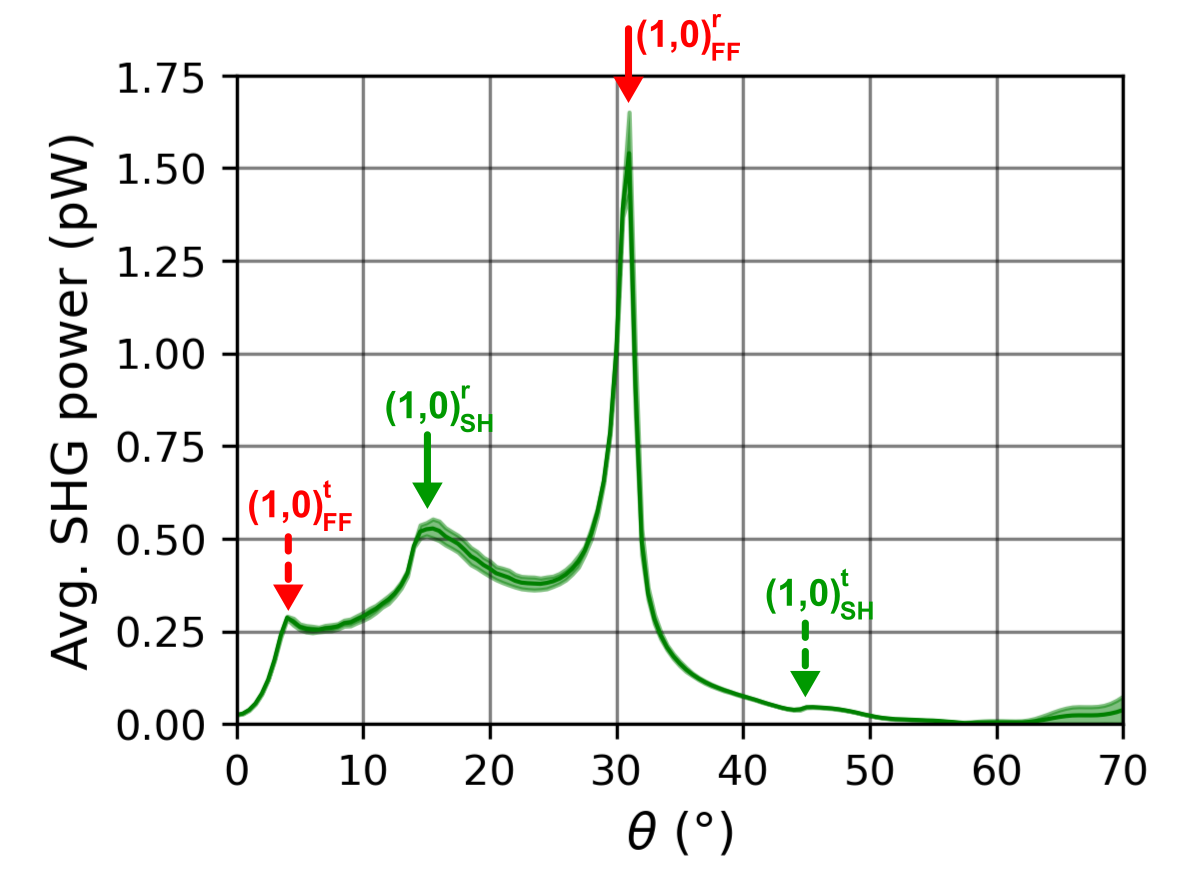}
	\caption{
		Average SHG power versus the incidence angle excited with TM-polarized light and $P_{FF} = \SI{50}{\milli\watt}$.
		RA condition $\theta_{RA}$ \eqref{eq:theta_RA} is at:
		$(1,0)^t_{FF}=\SI{3.7}{\degree}$,
		$(1,0)^r_{SH}=\SI{14.0}{\degree}$,
		$(1,0)^r_{FF}=\SI{31.0}{\degree}$,
        $(1,0)^t_{SH}=\SI{44.7}{\degree}$.}
	\label{Fig:SHG}
\end{figure}

Figure~\ref{Fig:SHG} shows the angle-dependent SHG response for TM/p-polarized excitation for a fixed average pump power of $P_{FF} = \SI{50}{\milli\watt}$.
While the centrosymmetry crystal lattice structure of gold is broken at the metal-dielectric interface \cite{Butet2015}, SHG emission in the far-field is still prohibited due to the spatial symmetry of the electric field distribution at normal incidence ($\theta = \SI{0}{\degree}$) \cite{Noor2020}.
However, oblique incidence breaks this spatial symmetry and allows SHG from the structure \cite{Ciraci2012}.
This also means that the nonlinear response does not change when the direction of the incoming wavevector is flipped.
This results in a symmetric SHG signal for positive and negative $\theta$s with respect to $\theta=\SI{0}{\degree}$.
We therefore determined the average SH signal for each absolute angle, while their difference is indicated as a filled area in Fig.~\ref{Fig:SHG}.
The difference between positive and negative $\theta$ is minimal.

Clearly, the SLRs have a significant impact on the frequency conversion process.
By varying $\theta$, different SLR of the metasurface are excited, resulting in three pronouced peaks in the SHG signal.
Near these angles, the RA condition for the transmission grating case $(1,0)^t_{FF}=\SI{3.7}{\degree}$, the reflection grating case $(1,0)^r_{FF}=\SI{31.0}{\degree}$, and the reflection grating case of the emission wavelength $(1,0)^r_{SH}=\SI{14.0}{\degree}$ is fulfilled (Eq. \eqref{eq:theta_RA}).
The excitation of the SLR at $(1,0)^r_{FF}$ produces the largest SHG signal.

\subsection{SHG input-output measurement}
\begin{figure}[tbp]
	\centering
	\includegraphics[width=.4\textwidth]{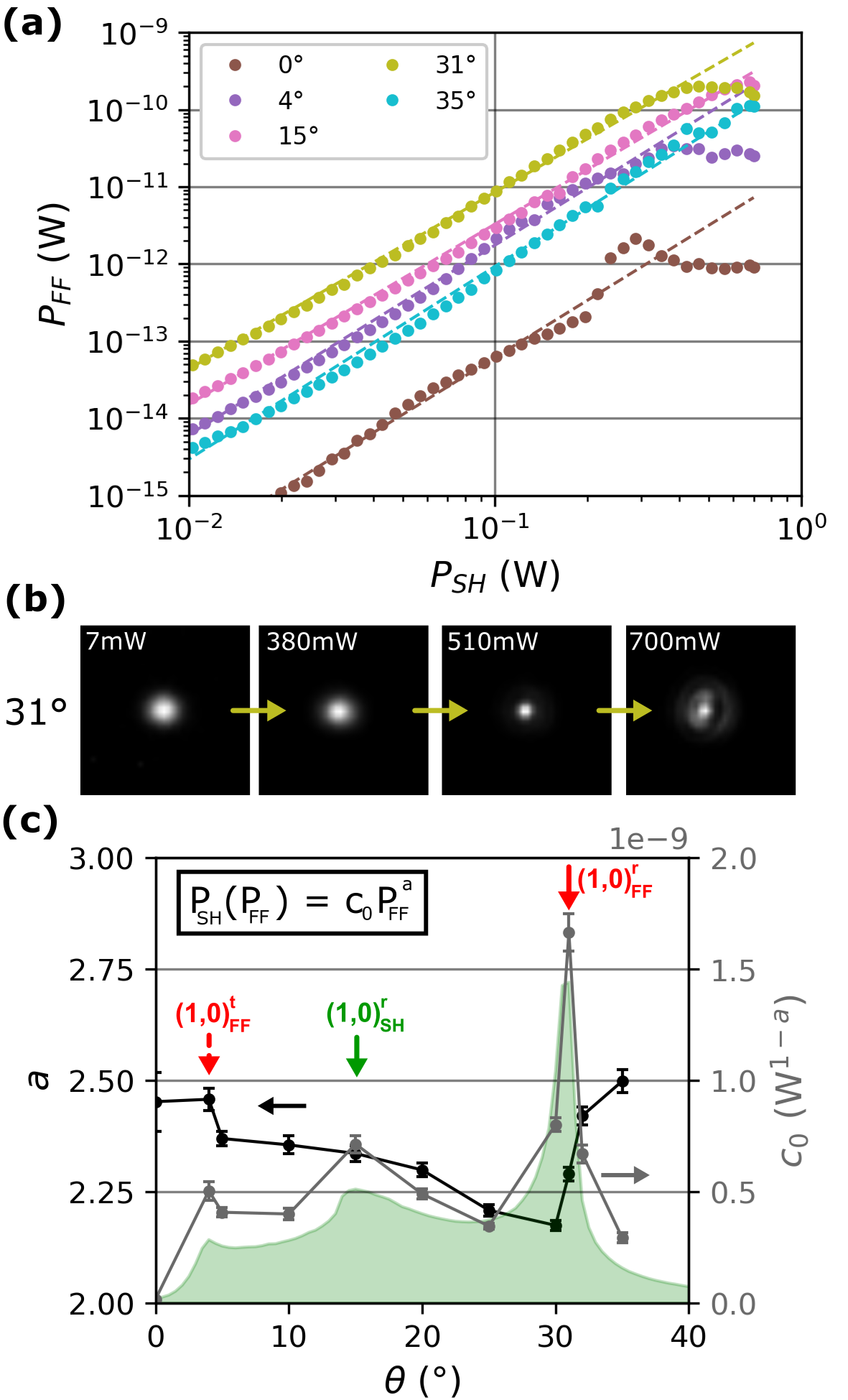}
	\caption{
		(a) Average SHG power depending on the incidence fluence $F_0$ for specific $\theta$. 
		Dashed lines show the fitted curve according to a power function $P_{SH} = c_0 P_{FF}^a$.
        (b) Recorded SHG beams at different pump powers for $\theta=\SI{31}{\degree}$.
		(c) Fitting parameter $c_0$ and $a$ depending on $\theta$. 
        The filled region represents the SHG signal, as shown in Fig.~\ref{Fig:SHG}.}
	\label{Fig:SHG_slope}
\end{figure}

For specific angles, the average optical pump power was gradually increased up to $P_{FF}=\SI{700}{\milli\watt}$, while simultaneously recording the emitted SHG power ($P_{SH}$) and the SHG beam profile.
At normal incidence, this maximum pump power corresponds to a peak incident fluence of $F_0 = \SI{58.9\pm 0.6}{\milli\joule\per\square\centi\meter}$, defined as the maximum fluence at the center of the Gaussian beam profile.
The position on the metasurface was shifted by around \SI{250}{\micro\meter} between each measurement with a different $\theta$, to ensure an unexposed metasurface at the beginning of each experiment.
Figure~\ref{Fig:SHG_slope}(a) shows the input-output response for a selected set of $\theta$ values.
Up to a certain value, the output power ramps up nonlinearly with the incident fluence or average pump power.
Above this critical fluence, the SHG signal saturates and the recorded Gaussian-shaped SH beam becomes distorted, see Fig.~\ref{Fig:SHG_slope}(b).
The nonlinear response of the metasurface is permanently modified once the fluence has been raised above this threshold, suggesting that the fluence in the center of the beam exceeds the LIDT of the metasurface.
Increasing the pump power leads to a larger central area in which the threshold is exceeded.
In contrast, the surrounding undamaged regions of the metasurface continues to produce SHG, which explains why the net detected signal remains approximately constant.\\

The nonlinear relationship between the input and output power can be described by $P_{SH} = c_0 P_{FF}^a$.
This power-law function is fitted in the region below the critical fluence and is represented by dashed lines in Fig.~\ref{Fig:SHG_slope}(a).
The two fitted parameters $c_0$ and $a$ are shown in Fig.~\ref{Fig:SHG_slope}(c).
For a second-order nonlinear process, such as the SHG, a slope of \SI{2}{} is expected.
Here, the slope $a$ is always greater than \SI{2}{}.
As a calibration measurement, the SHG from a nonlinear crystal and the glass substrate surface exhibits the expected quadratic power dependence with a slope of \SI{2.01\pm0.02}, confirming the reliability of the experimental setup.
Slightly steeper slopes than \SI{2}{} have previously been reported for nonlinear processes in metallic nanostructures \cite{Park2012, Stolt2022, Beer2022}.
Ultrafast excitation of the electron system can modify both the complex dielectric function \cite{Sun1994} and the second-order surface susceptibility of gold \cite{Hohlfeld1996, Guo2001}.
These transient changes can lead to spectral shifts and broadening of the plasmonic resonance, followed by recovery as the excited electrons relax \cite{Brown2017}.
Consequently, the plasmonic field enhancement can depend on the absorbed energy, potentially modifying the power dependence of the SHG signal \cite{Schirato2023}.
Multiphoton photoluminescence from gold nanoparticles \cite{Wang2023} can also increase the slope, but such a broad-band signal was not visible in the emission spectrum of this metasurfaces \cite{Beer2022}.
Consequently, changes in the plasmonic field enhancement and the intrinsic nonlinear response may contribute to the observed superquadratic power dependence of SHG.

The slope $a$ depends on $\theta$, revealing different nonlinear behavior depending on which SLR is excited.
In fact, a sudden change in the slope is visible at the SLR conditions for the pump wavelength.
At $(1,0)^t_{FF}$ the slope drops and at $(1,0)^r_{FF}$ it rises abruptly.
However, no change is observed for the RA condition of the second-harmonic $(1,0)^r_{SH}$.
This indicates that the modification of the nonlinear scaling is primarily governed by resonances for the fundamental wavelength.

The prefactor of the fitted function $c_0$ is related to the SHG efficiency $\eta^{(2)}_{SHG}=P_{SH}/P_{FF}^2$.
Therefore, its angular dependence is expected to correspond to the angle-dependent SHG signal measured at constant pump power in Fig.~\ref{Fig:SHG}.
For comparison, this SHG signal is reproduced as the filled region in Fig.~\ref{Fig:SHG_slope}(c). 
Both curves show good agreement.
However, due to the angle-dependency of the slope $a$, small deviations are expected.
The SHG efficiency related parameter $c_0$ increases when the SLR conditions are excited, with $(1,0)^r_{FF}$ exhibiting the highest conversion efficiency. \\

\section{Post-mortem analysis}
\begin{figure}[htbp]
	\centering
	\includegraphics[width=.5\textwidth]{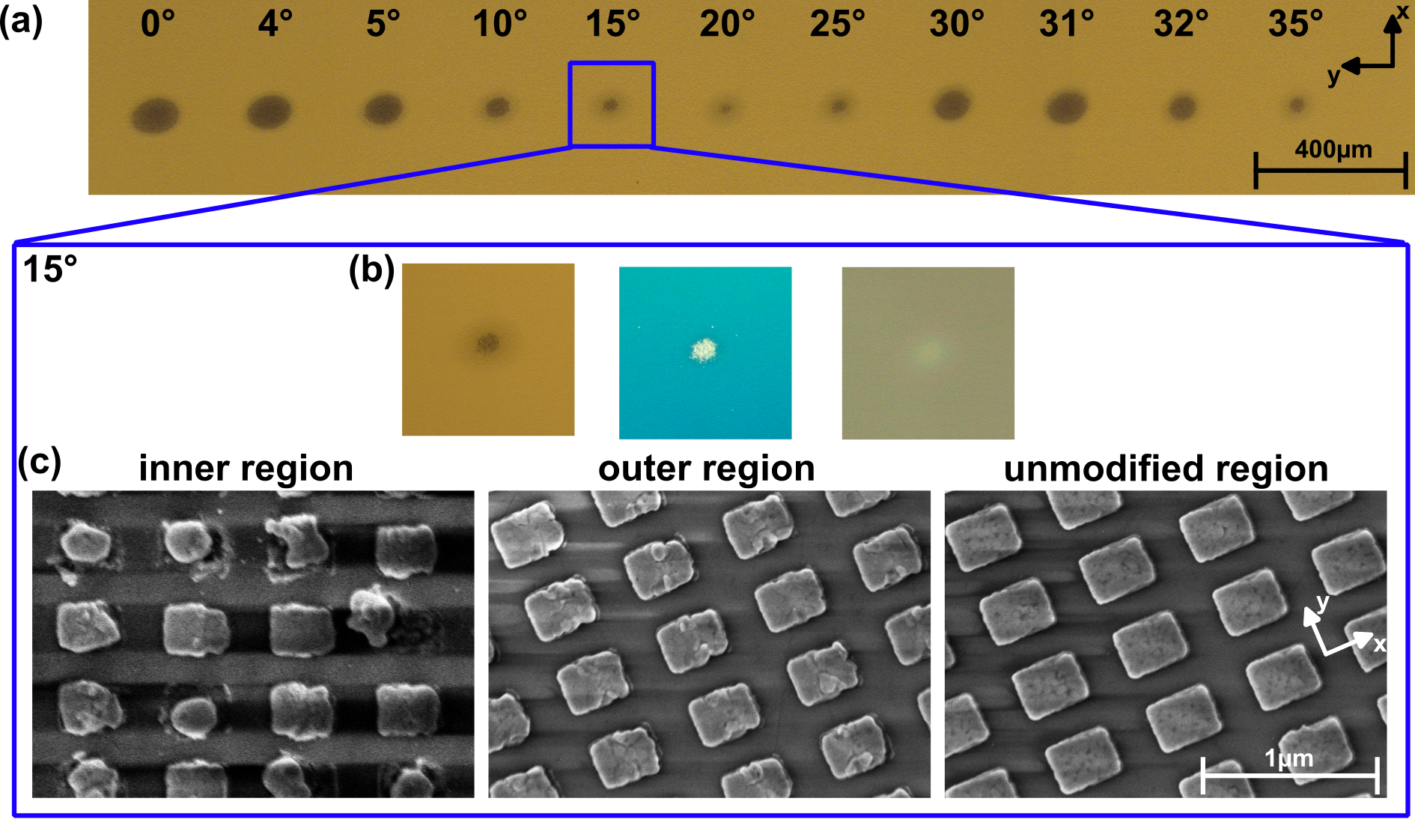}
	\caption{
		(a) Reflective bright-field optical microscope image of the metasurface after continuous irradiation with $F_0=\SI{59}{\milli\joule\per\square\centi\meter}, \Delta t=\SI{270}{\femto\second}, \lambda=\SI{1030}{\nano\meter}$ at various angles of incidence.
		(b) Optical microscope images for $\theta = \SI{15}{\degree}$ in different imaging modes, from left to right: reflective bright-field, reflective dark-field, and transmission bright-field.
		(c) SEM images showing different regions of the modified metasurface.
	}
	\label{Fig:LIDT_NB_SEM}
\end{figure}

After the input-output experiment, the sample was inspected under an optical microscope, see Figure~\ref{Fig:LIDT_NB_SEM}(a).
For all $\theta$, the maximum incident peak fluence caused a clearly visible modification of the metasurface.
The modifications consist of an inner region, where the reflection is reduced while transmission and scattering are increased compared to the unmodified region, and an outer region where the reflection and the color of the metasurfaces are slightly changed, shown in Fig.~\ref{Fig:LIDT_NB_SEM}(b).
The size of the inner region seems related to the SLR of the pump wavelength.
At the RA conditions of the pump wavelength $(1,0)_{FF}^t=\SI{3.7}{\degree}$ and $(1,0)_{FF}^r=\SI{31.0}{\degree}$ the inner region is enlarged.
On the other hand, the size of the outer region changes little with $\theta$.

\begin{figure*}[!t]
	\centering
	\includegraphics[width=\textwidth]{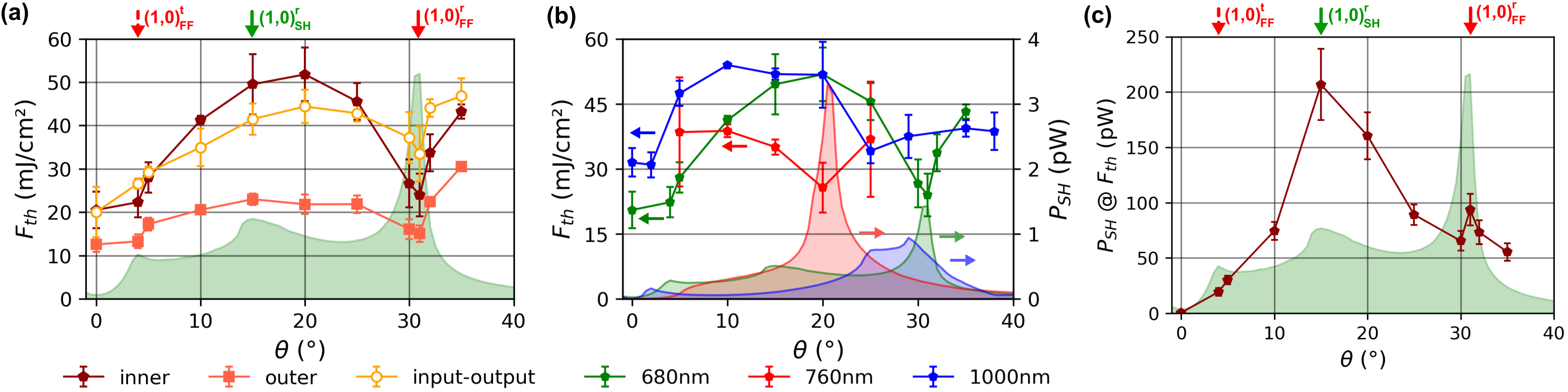}
	\caption{
		(a) Comparison of the threshold fluence obtained with two methods.
        The $F_{th}$ values of the 'outer' and 'inner' are determined using the D\textsuperscript{2}-method based on the diameters of the outer and the inner modification regions, respectively.
        The 'input-output' is the $F_{th}$ when the SH starts to look distorted.
        (b) The LIDT threshold fluence of the inner modification region (solid lines) and the detected SHG signal (filled region) for three periodicities $\Lambda_x$ of the gold NB metasurfaces.
        (c) Maximum achievable SHG signal when exciting the metasurface at the threshold surface fluence $F_{th}$.
        The filled region represents the detected SHG signal with an average pump power of $P_{FF} = \SI{50}{\milli\watt}$ ($F_0 = \SI{4.2}{\milli\joule\per\square\centi\meter}$).
	}
	\label{Fig:LIDT_NB}
\end{figure*}

The scanning electron microscope (SEM) images in Fig.~\ref{Fig:LIDT_NB_SEM}(c) reveal the structural change of the NBs across different regions: in the inner region, the NBs are randomly reshaped.
In some cases, the NBs are transformed into nearly round nanoparticles \cite{Link1999,Link2000,Wang2011}.
The removal of gold leads to the observed reduction in reflection and increase in transmission in this region (Fig.~\ref{Fig:LIDT_NB_SEM}(b)).
In the outer region, only one side of the NBs is altered compared to the unmodified metasurface.
This asymmetric modification of the NBs could be related to a $\theta$ induced near-field asymmetry.\\

From the maximum incident pump fluence of the Gaussian shaped pump beam and the measured diameter of the modification regions, 
the threshold fluence for the inner and outer regions is determined using the D\textsuperscript{2}-method \cite{Liu1982}:
\begin{align}
	F_{th} = F_{0}\cos\theta \cdot e^{-\dfrac{\cos\theta d_{th_x}d_{th_y}}{2\omega_0^2} }. \label{eq:F_th}
\end{align}
This analysis accounts for the incident angle $\theta$ dependent elongation of the illumination area on the metasurface, which causes a reduction of the effective fluence and an increase of the damage diameter along the $x$-direction ($d_{th_x}$), whereas the diameter in $y$-direction ($d_{th_y}$) should remain unaffected.
Figure~\ref{Fig:LIDT_NB}(a) compares the threshold fluence obtained from the post-mortem analysis of the inner and outer region using an optical microscope.
In addition, the threshold fluence where the beam profile starts to change during the input-output measurements (Fig.~\ref{Fig:SHG_slope}) is shown.
Clearly, the modification of the inner region is correlated with the distortion of the SHG beam shape and the saturation of the SHG signal.
The random reshaping of the gold NBs limits the achievable SHG signal.
The slightly altered NB shape in the outer region 
appears at approximately half the effective fluence than that of the inner region with no detectable influence on the input-output behavior and SHG beam shape.

It is clear that the threshold fluence $F_{th}$ depends on the incidence angle (Fig.~\ref{Fig:LIDT_NB}(a)).
Particularly noticeable is a dip in the threshold at the RA of $(1,0)^r_{FF}$, where the conversion efficiency is highest.
The $F_{th}$ is reduced when the SLR of the pump wavelength is excited, for both $(1,0)^t_{FF}$ and $(1,0)^r_{FF}$.
However, such a reduction is not observed for the SLR at the emission wavelength. \\

The fact that the SLR at the pump wavelength has a particularly important influence on the SHG process and the damage threshold can also be observed for other NBs periodicities.
Figure~\ref{Fig:LIDT_NB}(b) compares the LIDT threshold fluence of the inner modification region and the detected SHG signal measured at a fixed pump power for $\Lambda_x\in \{\SI{680}{\nano\meter}, \SI{760}{\nano\meter}, \SI{1000}{\nano\meter}\}$.
The other dimensions of the NBs remains as stated in Fig.~\ref{Fig:SampleDesign}, so the LSPR position is constant.
The SHG is enhanced when the SLR of the pump or the emission wavelength is excited \cite{Beer2022}.
The damage threshold is reduced when the SLR of the pump wavelength is excited, regardless of  the NBs periodicities.
However, this statement does not apply to the SLR at the SH.
From Fig.~\ref{Fig:LIDT_NB}(b), an average damage threshold of roughly \SI{30\pm15}{\milli\joule\per\square\centi\meter} can be estimated for he different metasurface periodicities.
For comparison, the LIDT of a \SI{50}{\nano\meter} thick gold thin film on fused silica substrate was determined using the same method under similar illumination conditions.
The determined threshold fluence of $F_0=\SI[separate-uncertainty=true]{143 \pm 17}{\milli\joule\per\square\centi\meter}$ is much larger than that of the metasurfaces.

While exciting the SLR related to $(1,0)^r_{FF}$ leads to the highest SHG efficiency, it also decreases the susceptibility to withstand laser-induced damage.
Since the SHG signal increases nonlinearly with pump intensity, the maximum achievable SHG output depends on both the conversion efficiency and the LIDT.
Figure~\ref{Fig:LIDT_NB}(c) shows the maximum achievable SHG power of the metasurface. 
This value is calculated by evaluating the fitted SHG power dependence from Fig.~\ref{Fig:SHG_slope}(c) at the excitation fluence corresponding to the measured LIDT of the inner region (Fig.~\ref{Fig:LIDT_NB}(a)).
The highest achievable SHG signal is obtained when an SLR is resonant with the second-harmonic wavelength $(1,0)^r_{SH}$. 
In contrast, resonant excitation at the fundamental wavelength of $(1,0)^r_{FF}$ and $(1,0)^t_{FF}$ provides only a minor increase in the maximum SHG output compared with the non-resonant case.
Thus, optimizing the SLR at the emission wavelength can be more effective for maximizing the achievable damage-free SHG output than optimizing it at the pump wavelength.

\section{Simulated linear optical properties}
Excitation of the SLR at the pump wavelength reduces the LIDT of the metasurface.
Finite difference time domain (FDTD) simulations are performed to relate this reduction to the linear optical response at the pump wavelength and to identify which simulated optical quantity correlates most strongly with the experimentally observed LIDT. 
In particular, the overall near-field enhancement, the locally enhanced peak field \cite{Plech2006}, and the optical absorption are considered. 
Distinguishing between these quantities is challenging because they are intrinsically interrelated.
Since the SLR for the SH appears to have no influence on the LIDT, the behavior at $\lambda_{SH} = \SI{515}{\nano\meter}$ will not be considered further.

\begin{figure}[htbp]
	\centering
	\includegraphics[width=.3\textwidth]{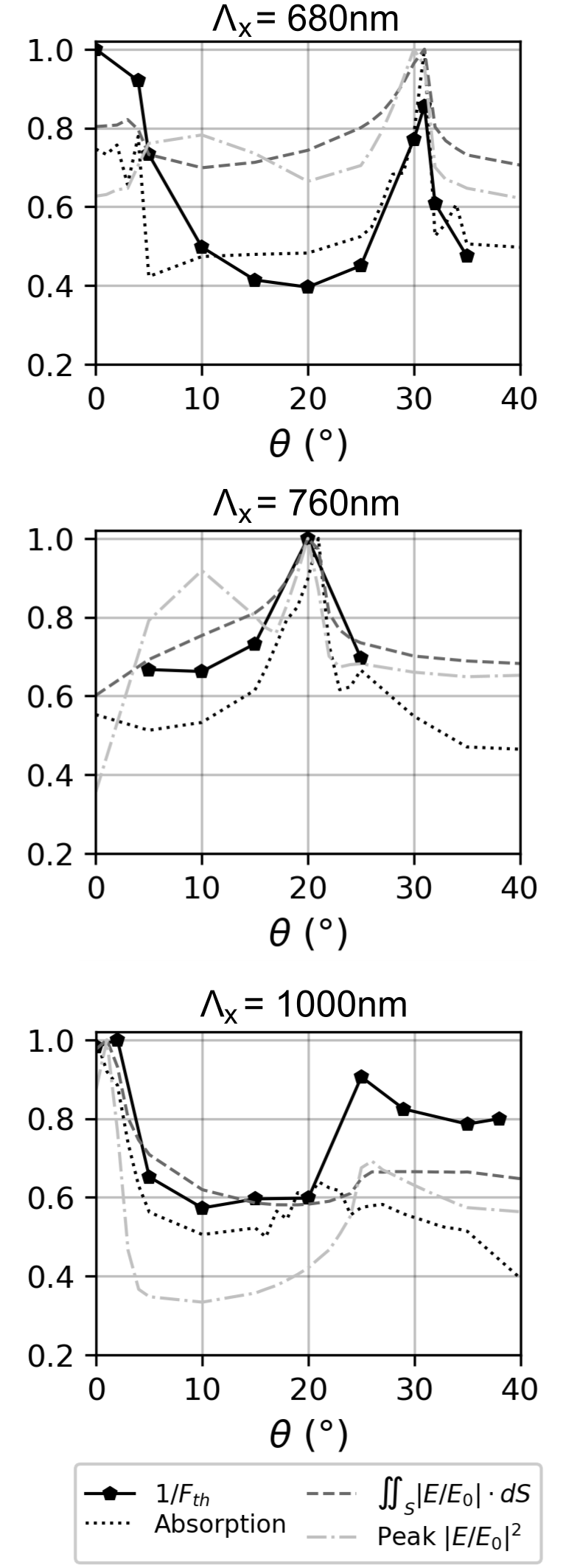}
	\caption{
		Reciprocal of the angle-dependent threshold fluence determined using D\textsuperscript{2}-method from the inner region $F_{th}$ compared with simulated linear response for the three metasurfaces, ($\Lambda_x=\SI{680}{\nano\meter}$, $\SI{760}{\nano\meter}$, and $\SI{1000}{\nano\meter}$, separately).
		The simulations include the normalized angle-dependent absorption, total integrated electric field, and the local peak intensity \SI{2}{\nano\meter} around the nanobar for $\lambda_{FF} = \SI{1030}{\nano\meter}$.
	}
	\label{Fig:LIDT_FDTD}
\end{figure}

In Fig.~\ref{Fig:LIDT_FDTD} the simulated angle-dependent quantities are compared with the reciprocal of the measured damage threshold obtained from the inner modification region (from Fig.~\ref{Fig:LIDT_NB}(b)).
For better comparison, all curves are normalized to their respective maximum values.
At first glance, the evolution of all curves appears similar, as all exhibit a peak at the SLR related to $(1,0)^r_{FF}$.
Among the simulated quantities, the absorption shows the closest qualitative agreement with the inverse LIDT.
However, when comparing the different periodicities, neither the optical absorption of the metasurface, nor the total field enhancement, nor peak intensity near the gold nanobar fully reproduces the observed LIDT behavior.
Thus, none of these quantities alone provides an unambiguous explanation of the damage mechanism.
The laser-induced modification of plasmonic nanostructures is a complex process governed by several factors, including optical absorption, thermal transport, heat distribution, and material-specific melting and diffusion processes \cite{Baffou2013, Ihm2019}.
Unlike the substrate damage reported by \citeauthor{Nagy2020}~\cite{Nagy2020}, no visible substrate modifications are observed in the SEM images in Fig.~\ref{Fig:LIDT_NB_SEM}(c), suggesting that substrate damage is unlikely to be the dominant mechanism.
Furthermore, the relative importance of different damage mechanisms can depend on the excitation conditions, such as repetition rate, pulse duration, and wavelength, as well as on the material composition and nanostructure geometry.
Based on the present results, optical absorption at the excitation wavelength appears to be a suitable first-order indicator of the angular dependence of the LIDT in the investigated plasmonic metasurfaces.\\

\section{Conclusion}
We presented a comprehensive study of the laser induced-damage threshold in periodic plasmonic gold nanobar metasurfaces and its direct impact on the SHG.
Through the combination of angle-resolved SHG measurements and a detailed LIDT analysis using the D\textsuperscript{2}-method, a clear relationship between SLR excitation, SHG efficiency, and the damage threshold was demonstrated.
Excitation of the SLRs at the pump wavelength significantly enhances SHG.
We show that it simultaneously induces an abrupt change in the nonlinear scaling of the SHG process and substantially reduces the LIDT.
In contrast, excitation of SLRs at the SH wavelength can increase SHG efficiency without noticeably reducing the damage threshold, allowing the SHG signal to be maximized.
This finding provides a promising strategy for optimizing frequency conversion in periodic plasmonic metasurfaces.

\begin{acknowledgments}
This work is funded by Deutsche Forschungsgemeinschaft (CRC 1375 NOA, 398816777).
The sample fabrication was carried out by the microstructure technology team at IAP Jena. 
The authors would like to thank them for providing the fabrication facilities, carrying out processes, and providing support.
\end{acknowledgments}

\bibliography{references}

\clearpage

\pagebreak
\widetext
\begin{center}
\textbf{\large Supporting Information: \linebreak "Laser induced damage threshold of periodic plasmonic metasurfaces"}
\end{center}

\setcounter{equation}{0}
\setcounter{figure}{0}
\setcounter{table}{0}
\setcounter{page}{1}
\setcounter{section}{0}
\makeatletter
\renewcommand{\theequation}{S\arabic{equation}}
\renewcommand{\thefigure}{S\arabic{figure}}
\renewcommand{\bibnumfmt}[1]{[S#1]}

\section*{Simulated normal incidence spectra}
Finite difference time domain (FDTD) simulations are performed using Lumerical FDTD by Ansys.
For investigating the linear spectral response of the periodic metasurface, power monitors are placed sufficiently far above and below the nanobars (NBs).
Due to the small bandwidth, Bloch source and Bloch-periodic boundaries are used.
These monitors record the total spectral power density passing through the monitor.
However, this total power includes contributions from all diffraction orders, 
whereas experimentally only certain diffraction orders are detected.
Therefore, using the built-in grating projection analysis, the optical power propagating in a certain direction is extracted from the power monitor data.
These monitors can also be used to calculate other optical parameters that are difficult to determine experimentally.
The absorption is obtained from the power conservation law and corresponds to the fraction of the incident power that is neither transmitted nor reflected.
The difference between the total power and the power in the 0\textsuperscript{th} diffraction order is the grating scattering, containing the diffracted and scattered light of the metasurface.\\

\begin{figure}[htbp]
	\centering
	\includegraphics[width=.9\textwidth]{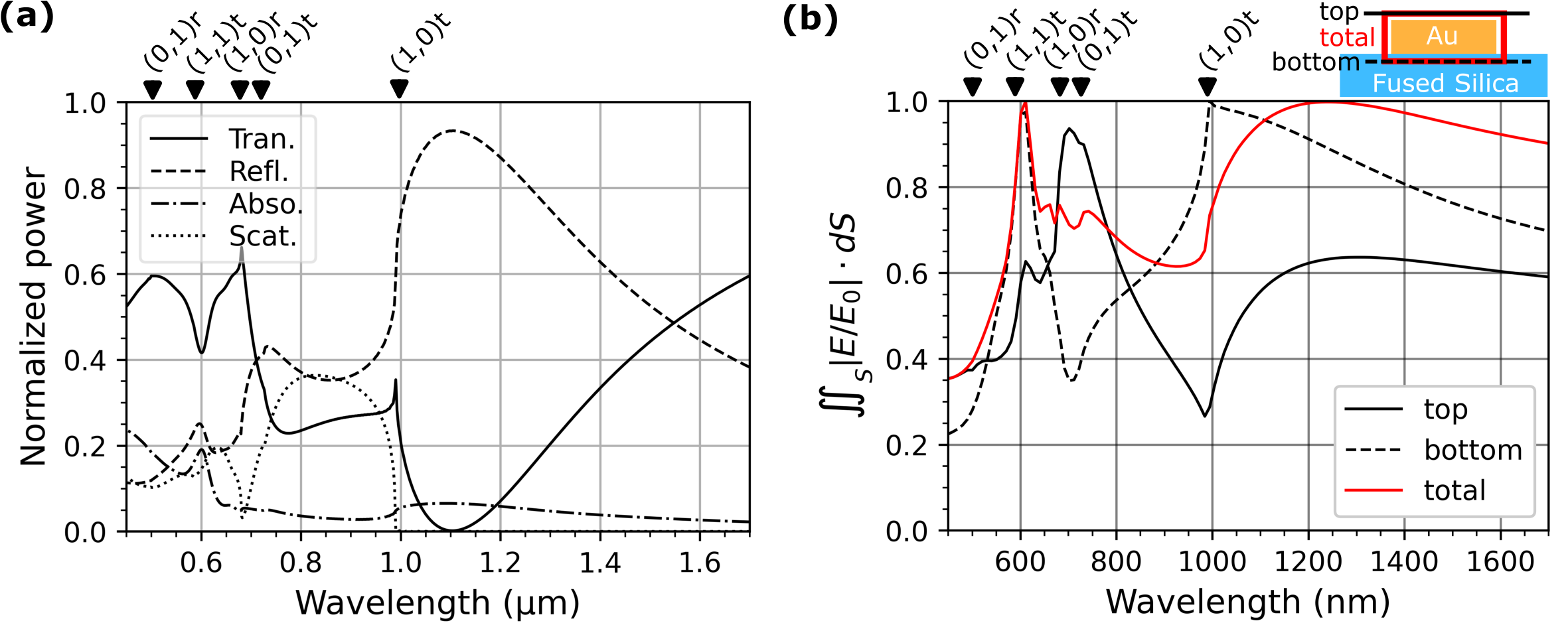}
	\caption{
		FDTD simulated linear spectral response for $\Lambda_x=\SI{680}{\nano\meter}$ at normal incidence.
        (a) Linear spectrum.
        (b) Integrated field enhancement at the surface \SI{2}{\nano\meter} below the fused silica interface (bottom), \SI{2}{\nano\meter} above the NB (top) as well as \SI{2}{\nano\meter} around the entire NB surface (total).
		The curves of the integrated electric field are normalized to their maximum values. 
		RA conditions $\theta_{RA}$ occur at:
		$(1,0)^t=\SI{986}{\nano\meter}$,
		$(0,1)^t=\SI{727}{\nano\meter}$,
		$(1,0)^r=\SI{680}{\nano\meter}$,
		$(1,1)^t=\SI{589}{\nano\meter}$,
		$(0,1)^r=\SI{500}{\nano\meter}$,
		$(2,0)^t=\SI{497}{\nano\meter}$.}
	\label{Fig:FDTD_spectra_normal}
\end{figure}

The simulated linear spectra at normal incidence for $\Lambda_x=\SI{680}{\nano\meter}$ is shown in Fig.~\ref{Fig:FDTD_spectra_normal}(a).
The localized surface plasmon resonance (LSPR) of the metasurface appears near $\lambda \approx \SI{1100}{\nano\meter}$, with an increased reflection but decreased transmission.
In the absorption spectrum, the LSPR of the metasurface appears as a broad absorption band with around \SI{7}{\percent} peak absorbance.
In contrast, the multiple surface lattice resonances (SLRs) appear as narrow spectral features.
Especially noticeable in the transmission and grating scattering spectra.
In the visible range the absorption increases strongly towards shorter wavelengths due to interband transitions in gold, which lead to strong intrinsic material losses.
Consequently, the absorption at $\lambda_{SH} = \SI{515}{\nano\meter}$ is higher than at the near-infrared excitation wavelength $\lambda_{FF} = \SI{1030}{\nano\meter}$, close to the LSPR.

Figure~\ref{Fig:FDTD_spectra_normal}(b) shows normalized integrated electric fields at a distance of \SI{2}{\nano\meter} from the NB.
A distinction is made between surfaces \SI{2}{\nano\meter} above the NB (top, solid line), \SI{2}{\nano\meter} below the NB (bottom, dotted line), and \SI{2}{\nano\meter} around the entire NB (total, solid red line).
The electric field is enhanced on the top side of the NB when the RA condition in the reflection grating case $(1,0)^r$ is excited.
At the same time, the field enhancement on the bottom of the NB is typically weakened.
In contrast, the field is enhanced on the bottom side of the NB when the RA condition in the transmission grating case $(1,0)^t$ is excited.
When integrating around the entire NB, the SLR do not produce a pronounced peak but appears only as an anomaly.
Exciting the SLR mainly redistributes and localizes the field, but does not significantly increase the total integrated field around the nanostructure.\\

\section*{Simulated angle-dependent response}
\begin{figure}[htb]
	\centering
	\includegraphics[width=0.5\textwidth]{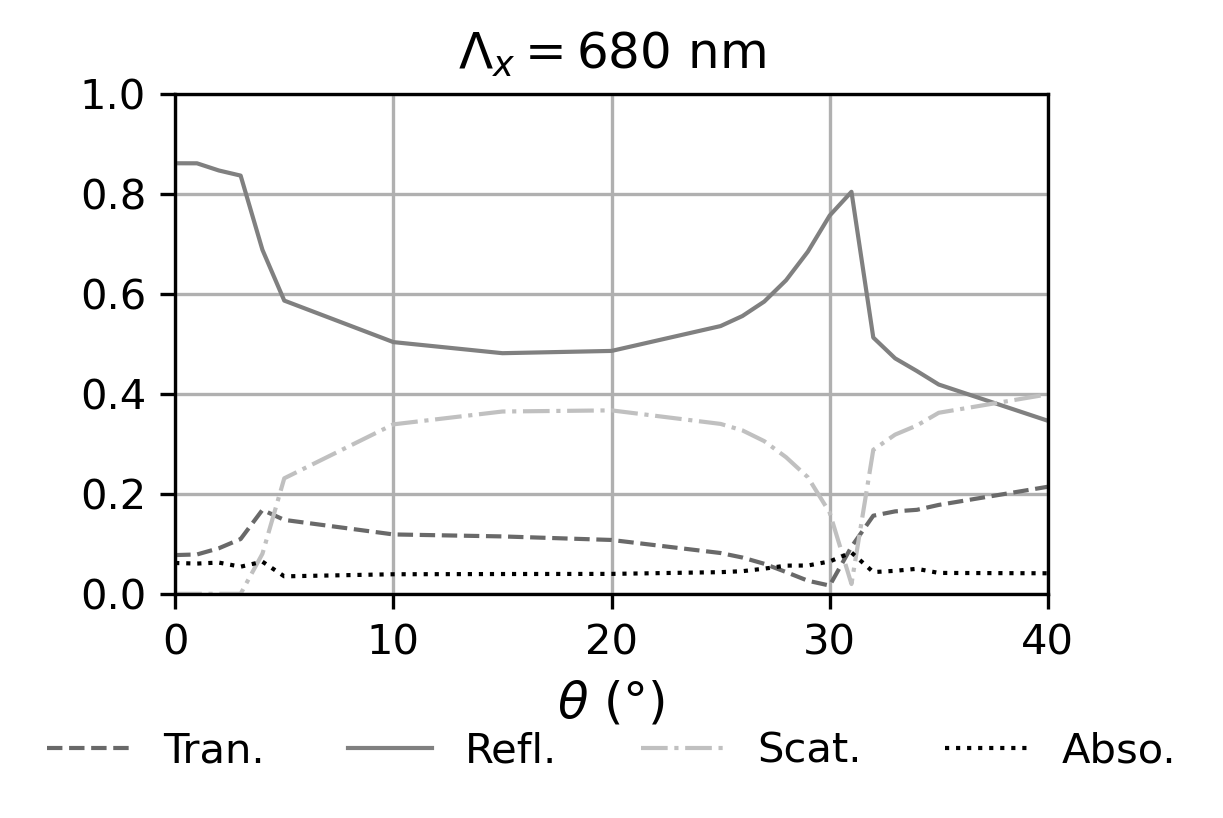}
	\caption{
		FDTD simulated angle-dependent linear response for $\lambda_{FF}=\SI{1030}{\nano\meter}$.
	}
	\label{Fig:NB_Px=680nm_FDTD_AOIscan}
\end{figure}

Figure~\ref{Fig:NB_Px=680nm_FDTD_AOIscan} shows the FDTD simulated linear response for $\lambda_{FF}=\SI{1030}{\nano\meter}$ depending on the angle of incidence.
When exciting the SLR related to $(1,0)^r_{FF}=\SI{31.0}{\degree}$, the reflectance and absorbance are increased but the grating-scattering is reduced.
An anomaly is visible for the angle-dependen transmittance.
On the other hand, for $(1,0)^t_{FF}\approx\SI{3.7}{\degree}$ the transmittance and absorbance increased while an anomaly is visible for the reflectance and grating-scattering.

\section*{Comparison of different periodicities}
\begin{figure}[htb]
	\centering
	\includegraphics[width=0.9\textwidth]{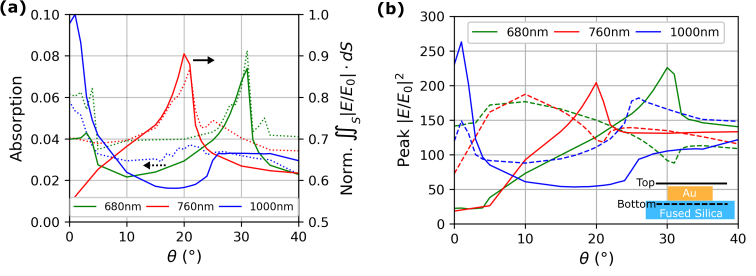}
	\caption{
		FDTD simulated linear response for the three NB metasurfaces with $\Lambda_x=\SI{680}{\nano\meter}, \SI{760}{\nano\meter}, \SI{1000}{\nano\meter}$.
		(a) Corresponding angle-dependent absorption for $\lambda_{FF} = \SI{1030}{\nano\meter}$ (solid) and total integrated electric field around the NB (dotted) for the three periodicities.
		(b) Angle-dependent peak intensity at bottom surface, \SI{2}{\nano\meter} below the fused silica interface (dashed), and at the top surface (solid), \SI{2}{\nano\meter} above the NB for $\lambda_{FF} = \SI{1030}{\nano\meter}$.
	}
	\label{Fig:NB_FDTD}
\end{figure}

Figure~\ref{Fig:NB_FDTD}(a) shows the angle-dependent absorption for the three experimentally investigated NB metasurface periodicities for TM-polarized light at $\lambda_{FF} = \SI{1030}{\nano\meter}$, shown as dotted lines.
In addition, the angle-dependent integrated total electric field enhancement around the NB at a distance of \SI{2}{\nano\meter} from the gold surface is shown as solid lines.
The data are normalized to the maximum value among the three periodicities.
A high total electric field enhancement indicates that the incident light efficiently excites plasmonic modes of the nanoparticle,
implying strong induced charge oscillations \cite{Maier2007}.
Clearly, the normalized total field enhancement is related to the absorption of the metasurface.
Both quantities are intrinsically linked \cite{Maier2007}.
The peaks in the absorption and the total electric field enhancement of Fig.~\ref{Fig:NB_FDTD}(a) are related to the SLR \cite{Humphrey2014, Kravets2018}.
At this resonance, the incident energy couples efficiently into the collective mode but cannot easily leave the system via radiation.
The radiative decay rate is reduced \cite{Kravets2018}, which also appears as a clear dip in the grating-scattering spectrum (Fig.~\ref{Fig:NB_Px=680nm_FDTD_AOIscan}).
As a result, more optical energy is temporarily stored and dissipated within the metasurface, leading to an increased total field enhancement.
In contrast, the non-radiative losses are mainly determined by intrinsic material losses in the metal, so a larger fraction of the energy is dissipated as heat.
Exciting the SLR modifies the balance between radiative and non-radiative decay channels, thus increasing the absorption of the metasurface. 
\newpage

Plasmonic structures produce strong spatial field localization.
Even when the global fluence is below the nominal LIDT of the materials, the local peak intensity in the hot spots can trigger linear and nonlinear absorption processes.
Figure~\ref{Fig:NB_FDTD}(b) shows the angle-dependent peak intensity of the electric field enhancement.
A distinction is made between \SI{2}{\nano\meter} above (in air) and \SI{2}{\nano\meter} below the NB (in fused silica).
Clearly, the peak intensity exhibits an anomaly when the metasurface is excited at an angle corresponding to the SLR.
The electric field is enhanced on the top side of the NB when the RA condition in air $(1,0)^r$ is excited, see $\Lambda_x=\SI{680}{\nano\meter}$ $\theta = \SI{30}{\degree}$, $\Lambda_x=\SI{760}{\nano\meter}$ $\theta = \SI{20}{\degree}$, and $\Lambda_x=\SI{1000}{\nano\meter}$ $\theta = \SI{3}{\degree}$.
At the same time, the field enhancement on the bottom of the NB is typically weakened.
In contrast, the field exhibits an anomaly on the bottom side of the NB when the RA condition in fused silica $(1,0)^t$ is excited, see $\Lambda_x=\SI{680}{\nano\meter}$ $\theta = \SI{4}{\degree}$, $\Lambda_x=\SI{760}{\nano\meter}$ $\theta = \SI{6}{\degree}$, and $\Lambda_x=\SI{1000}{\nano\meter}$ $\theta = \SI{25}{\degree}$.

\end{document}